\documentclass[twocolumn,english,aps,superscriptaddress,prb]{revtex4-1}
\usepackage{graphicx} 
\usepackage{float}
\usepackage{tikz}
\usepackage{geometry}
\usepackage{amsmath}
\usepackage{physics}

\newcommand{\emptycircle}{
  \begin{tikzpicture}[baseline=-0.6ex]
    \draw (0,0) circle (0.1);
  \end{tikzpicture}
}
\newcommand{\fullcircle}{
  \begin{tikzpicture}[baseline=-0.6ex]
     \filldraw[fill=black] (0,0) circle (0.1);
  \end{tikzpicture}
}

\begin{document}
\title{Resolving chiral transitions in Rydberg arrays with quantum Kibble-Zurek mechanism and finite-time scaling}

\author{Jose Soto Garcia}

\affiliation{Kavli Institute of Nanoscience, Delft University of Technology, Lorentzweg 1, 2628 CJ Delft, The Netherlands}
\author{Natalia Chepiga}
\affiliation{Kavli Institute of Nanoscience, Delft University of Technology, Lorentzweg 1, 2628 CJ Delft, The Netherlands}


\begin{abstract}
The experimental realization of the quantum Kibble-Zurek mechanism in arrays of trapped Rydberg atoms has brought the problem of commensurate-incommensurate transition back into the focus of active research. Relying on equilibrium simulations of finite intervals, direct chiral transitions at the boundary of the period-3 and period-4 phases have been predicted. Here, we study how these chiral transitions can be diagnosed experimentally with critical dynamics. We demonstrate that chiral transitions can be distinguished from the floating phases by comparing Kibble-Zurek dynamics on arrays with different numbers of atoms. Furthermore, by sweeping in the opposite direction and keeping track of the order parameter, we identify the location of conformal points. Finally, combining forward and backward sweeps, we extract all critical exponents characterizing the transition.
\end{abstract}
\pacs{
75.10.Jm,75.10.Pq,75.40.Mg
}

\maketitle

{\bf Introduction.}
Understanding the nature of quantum phase transitions (QPTs) in strongly correlated low-dimensional systems is one of the biggest challenges in modern condensed matter physics\cite{sachdev1999quantum,giamarchi}. 
The development of conformal field theory (CFT)\cite{difrancesco,tsvelik} has led to the discovery of many fascinating critical phenomena with dynamical critical exponent $z=1$. The latter implies, in particular, a quantum-classical correspondence, which opens a way to study QPT with models of classical statistical mechanics. 

The construction of advanced entanglement-based numerical techniques such as tensor network algorithms \cite{dmrg1,dmrg3,schollwock2011density,PAECKEL2019167998} and recent progress in quantum simulating platforms have further stimulated the discovery of exotic critical phenomena, putting a focus onto quantum versions of phase transitions. The study of phase transitions beyond CFT, although extremely challenging, has been attracting a lot of attention in the past decades.   One of the most intriguing and debated problem is a possibility of a direct chiral transition out of crystalline period-3 phase predicted by Huse and Fisher\cite{huse1982domain} in the context of adsorbed monolayers\cite{Den_Nijs,SelkeExperiment,birgeneau,Selke1982,huse1984commensurate}.  

Recent experiments on Rydberg atoms have shed new light onto this originally classical problem, but now in the context of one-dimensional (1D) quantum chains  \cite{keesling2019quantum,scipost_chepiga,chepiga2019floating,maceira2022conformal,dalmonte,chepiga2021kibble,PhysRevB.98.205118,PhysRevResearch.3.023049,samajdar2018numerical,rader2019floating}. In these experiments, Rydberg atoms are trapped with optical tweezers at a well-controlled inter-atomic distance. Competition between the laser detuning, favouring atoms to be in Rydberg states, and strong van der Waals repulsions between them leads to a rich phase diagram dominated by lobes with integer periodicities $p=2,3,4,$...

Although quantum-classical correspondence does not hold for non-conformal criticality, many features of the transitions with $z\neq 1$ are qualitatively similar\cite{huse1982domain,chepiga2019floating,chepiga2021kibble,maceira2022conformal,PhysRevB.108.184425}. In particular, using Huse and Fisher's original criteria for chiral transitions and numerical simulations with state-of-the-art density matrix renormalization group algorithm\cite{dmrg1,dmrg3,schollwock2011density}, it has been shown that in Rydberg arrays, the transition out of $p=3$ phase changes its nature multiple times\cite{fendley,chepiga2019floating,dalmonte,maceira2022conformal}. At the point where chiral perturbations vanish, the transition is conformal in the three-state Potts universality class; away from this point but not too far from it the transition is direct in the Huse-Fisher universality class; further away the transition is a two-step process via conformal Kosterlitz-Thouless\cite{Kosterlitz_Thouless} and non-conformal Pokrovsky-Talapov\cite{Pokrovsky_Talapov} transitions, with a floating phase between the two. Quite surprisingly, numerical simulations revealed that the boundary of the $p=4$ phase undergoes a similar zoo of QPTs with a conformal Ashkin-Teller point, followed by the $p=4$ chiral transition, and then by the floating phase\cite{chepiga2021kibble,maceira2022conformal}.
   
Intervals of chiral transitions predicted numerically explain the experimental data obtained with Rydberg atoms. However, the numerical methods used to diagnose phase transitions differ greatly from the experimental techniques. Numerical simulations mainly rely on equilibrium physics and extracted critical exponents $\nu$ of the correlation length, $\bar{\beta}$ of the incommensurate wave vector, and $\alpha$ of the specific heat. 
At the same time, the experiments are performed out of equilibrium, employing quantum Kibble-Zurek (KZ) mechanism\cite{keesling2019quantum,zurek2005dynamics,dziarmaga2005dynamics}, which tracks the number of domain walls (kinks) formed upon sweeping through a phase transition as a function of sweep rate. This paper aims to fill the gap between the theoretical predictions for chiral transitions and experimentally accessible measurements to diagnose them. 

{\bf The Kibble-Zurek mechanism} describes the generation of topological defects during the constant-rate drive of a system through a second-order phase transition \cite{dziarmaga2010dynamics}.
The concept, introduced by Kibble to explain galaxy formation in the nascent universe  \cite{kibble1976topology}, was brought to condensed matter by Zurek\cite{zurek1985cosmological} and recently has been extended to  QPTs\cite{zurek2005dynamics,dziarmaga2005dynamics,damski2005simplest}. 
In the critical region, the correlation length $\xi \sim \left|g-g_c\right|^{-\nu}$ and the relaxation time $\tau \sim \left|g-g_c\right|^{-\nu z}$ diverge with the distance to the transition $\left|g-g_c\right|$. Sweeping through the transition in a non-adiabatic fashion with a given sweep rate $s$ creates a certain density of kinks (domain walls) $n_k$, that, in turn, determines the final correlation length $\hat \xi$ measured deep in the ordered phase. According to the KZ mechanism, the scaling of both quantities is universal and governed by the critical exponent $\mu= \nu/(1 + \nu z)$:
\begin{equation}
    n_k \sim \hat \xi^{-1} \sim s^{-\mu}.
\end{equation}

{\bf Finite-time scaling} (FTS) is the temporal analogue of the finite-size scaling performed in the critical region when the system size $L$ used in numerical (or quantum) simulations is smaller than the actual correlation length. FTS allows accessing the scaling of macroscopic quantities in the non-adiabatic region. For Rydberg arrays, such analysis has been performed only for the correlation length\cite{keesling2019quantum}. 

Inspired by the results for the chiral clock model\cite{huang2019nonequilibrium}, we perform FTS by sweeping from the ordered to the disordered phase - in the direction opposite to the conventional Kibble-Zurek drive. If the KZ mechanism tracks the formation of domain walls in the ordered phase, the backward sweeping captures the inertia of the order parameter - the faster one sweeps through the transitions, the deeper in the disordered phase is the point $g_0$ where the order parameter vanishes\cite{gong2010finite, huang2014kibble}:
 \begin{equation}
    |g_0(s)-g_c| \propto s^\frac{\mu}{\nu},
    \label{eq:fts1}
\end{equation}
At the critical point, the remaining order parameter $O_c$ scales with the sweep rate as\cite{gong2010finite, huang2014kibble}:
\begin{equation}
    O_c(s) \propto s^{\beta\frac{\mu}{\nu}}
    \label{eq:fts2}
\end{equation}
These complete the set of independent critical exponents, allowing to extract $\mu,\nu,\beta$ and $z$ and derive other critical exponents using the hyperscaling relation $2-\alpha=\nu (1+z)$.

{\bf The blockade model.}
An array of Rydberg atoms can  be described by the Hamiltonian of interacting hard-core bosons:
\begin{equation}
       H = \frac{\Omega}{2}\sum_i\left(d_i + d^\dagger_i\right)-\Delta\sum_i n_i
       + \sum_{i<j}V_{ij}n_in_j,
       \label{eq:tail}
   \end{equation}
where $\Omega$ is the Rabi frequency bringing an atom to the Rydberg state, $\Delta$ is laser detuning and $V_{ij} \propto 1/r^6$ is a van der Waals interaction. 

Owing to an extremely strong repulsion at short distances, we can write down an effective Hamiltonian with $r$-site blockade:
\begin{subequations}
\begin{equation}
    H = \sum_i -\frac{\Omega}{2}(d^\dagger_i+d_i) - \Delta n_i + V_{r+1}n_i n_{i + r + 1} \\
\label{eq:h_blockade}
 \end{equation}  
  \begin{equation}
    n_i\left( n_i - 1\right) = n_in_{i+1} = \dots = n_in_{i+r} = 0.
    \label{eq:blockade}
    \end{equation}
\end{subequations}
All operators in Eq.(\ref{eq:h_blockade}) are acting in the constrained Hilbert space defined by Eq.(\ref{eq:blockade}), which implies that two atoms cannot be excited within a certain distance $r$. The next-to-blockade interaction is encoded in the third term and the rest of the decaying interaction tail is set to zero. Various blockade ranges give access to different slices of the phase diagram\cite{chepiga2021kibble}: we will use $r=1$ blockade to probe chiral transition out of period-3 phase and $r=2$ for the period-4 case.
The advantage of the blockade model is three-fold: {\it i)} it provides a good approximation to the van der Waals potential\cite{chepiga2019floating,chepiga2021kibble,maceira2022conformal}; {\it ii)} tensor networks with explicitly implemented blockade have significantly lower computational costs\cite{scipost_chepiga,chepiga2021kibble}; {\it iii)} for these models, the location and extent of chiral transitions are known with a high accuracy\cite{chepiga2019floating,chepiga2021kibble}.

Numerical simulations were performed with the time-evolving block decimation algorithm (TEBD) with time step $\delta t = 0.1$ and maximal bond dimension $D = 400$ (see Supplemental materials\cite{SM} for further technical details).

{\bf Floating phase vs direct transition.} Our first goal is to resolve the direct chiral transition from the intermediate critical phase. Upon approaching the floating phase from the disordered one, the correlation length $\xi$ diverges stretch-exponentially due to Kosterlitz-Thouless transition\cite{Kosterlitz_Thouless}, leading to the effective exponent $\nu_\mathrm{eff}\rightarrow\infty$, and consequently $\mu_\mathrm{eff} \rightarrow 1$. Interestingly, this value is much larger than the one for the three-state Potts ($\mu=5/11$) or Ashkin-Teller ($0.4\leq \mu\leq 0.5$) transitions. But we rely on the fact that this scaling is only approached asymptotically and otherwise is affected by significant finite-size effects\cite{dziarmaga2014quench,gardas2017dynamics}.

\begin{figure}[h!]
\centering 
     \includegraphics[width=0.5\textwidth]{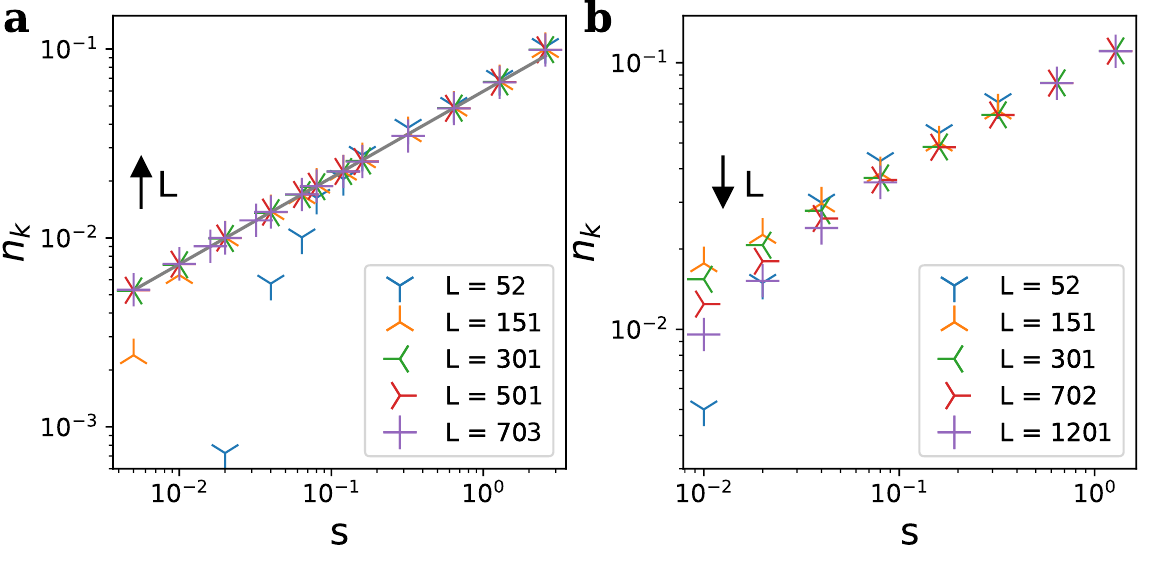}
        \caption{Scaling of the density of kinks $n_k$ with the sweep rate $s$ through (a) the direct transition; and (b) the floating phase. Arrows show systematic finite-size effects that can be used to resolve the two critical regimes. The grey line is a linear fit in a $\log-\log$ scale.}
        \label{fig:block_size_kt_chiral}    
\end{figure}

In Fig.\ref{fig:block_size_kt_chiral}(a), we present the scaling of the density of kinks\cite{SM} formed by sweeping through the direct transitions for various system sizes. For a given size $L$, there is a certain range of sweep rates $s$ where the scaling in $\log-\log$ plot is linear, in agreement with KZ mechanism. However, the curves turn down for small $s$, where the system dynamics approaches the adiabatic regime for a given $L$. This is similar to when one underestimates a correlation length when it is comparable to or exceeds the chain length. The window of the universal KZ scaling can be increased by increasing the length $L$. In other words, comparing the density of kinks at low sweep rates for several system sizes, we see that it grows with $L$\cite{SM}.

Across the floating phase, the finite-size effect is very different, as presented in Fig.\ref{fig:block_size_kt_chiral}(b): the larger the system size, the closer to the asymptotic limit, and therefore the steeper the scaling is. Focusing again on low sweep rates, we see a systematic and fast decrease in kinks density with $L$\cite{SM}. However, the limitation of measurable density of kinks for very small system size $L$ is still in place, as can be clearly seen for $L=52$. 

To summarize, for a set of sufficiently large system sizes, the density of kinks grows with the system size towards the universal KZ scaling when the system is driven through a direct transition and $n_k$ decays significantly when sweeping through the floating phase.

{\bf Conformal vs chiral transitions with KZ mechanism.} 
We have defined a protocol to distinguish the direct transition from the floating phase. Now let us see if with critical dynamics we could identify when the transition is conformal. For this purpose, we systematically extract the KZ critical exponent $\mu$ across various cuts into period-3 and period-4 phases. Our numerical results are summarized in Fig.\ref{fig:blockade_kz_distribution_overlap}. For the model with $r=1$ blockade, there is only one conformal critical point, and its location is known exactly\cite{fendley}. For the model with $r=2$ blockade, the location of the conformal point 
has been obtained numerically by tracking the commensurate line where chiral perturbations vanish\cite{chepiga2021kibble}. 

\begin{figure}[h!]
     \centering
     \includegraphics[width=\columnwidth]{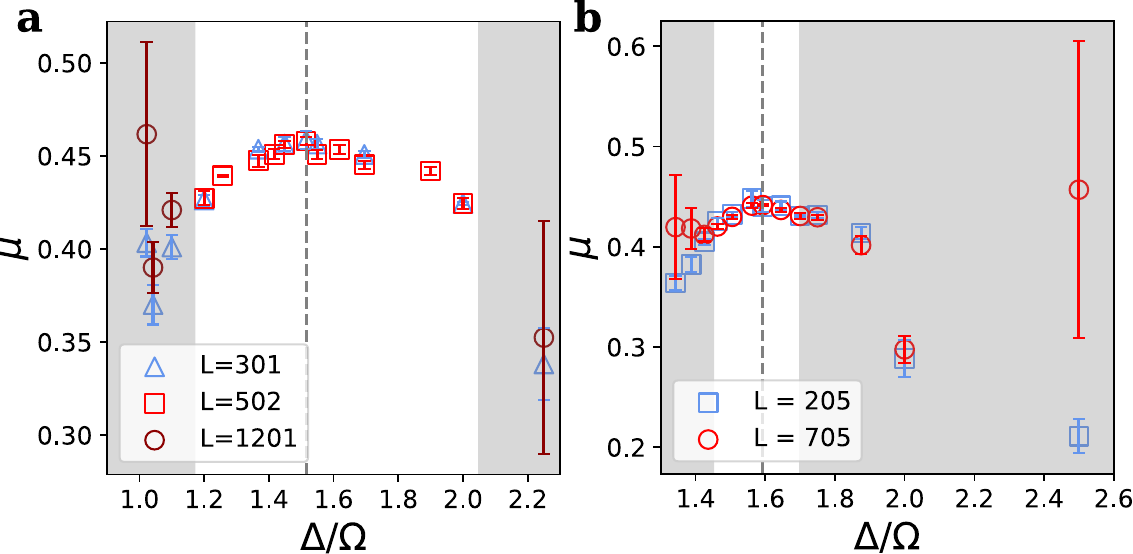}
    \caption{Critical exponent $\mu$ extracted with the Kibble-Zurek mechanism along various cuts through the transitions (a)  into the period-3 phase with $r=1$ blockade and (b) into the period-4 phase with $r=2$ blockade. The vertical dashed line indicates the location of the conformal point. The shaded region states for the floating phase. In this area, the Kibble-Zurek mechanism is not defined, and our results should be taken only as $\mu_\mathrm{eff}$.  At the conformal points (dashed lines), the extracted values of $\mu$ agree with theory predictions within $1\%$. Error bars reflect errors from the fit but do not show errors due to entanglement cutoff.}
    \label{fig:blockade_kz_distribution_overlap}
\end{figure}

For both transitions into period-3 and period-4 phases, the KZ critical exponent $\mu$ measured for a set of consecutive cuts has a dome shape, taking maximal values at the cuts that go through the conformal points. Numerically obtained values $\mu\approx0.458$ for the three-state Potts and $\mu\approx0.442$ for the Ashkin-Teller point (with $\nu\approx 0.78$\cite{chepiga2021kibble}) agree within $1\%$
with the CFT predictions. Away from conformal points, $\mu$ shows a slow decay. The dome-like shape we observe here is in excellent agreement with the experimental results\cite{keesling2019quantum}, and has been overlooked in the previous numerical simulations of dynamics hidden by large numerical errors\cite{keesling2019quantum}.

By looking at Fig.\ref{fig:blockade_kz_distribution_overlap} with marked conformal points, it is easy to associate them with maxima of $\mu$.   However, if the location of these points were unknown (as is often the case in experiments), then the problem would be quite challenging since the domes are relatively flat. As a solution, we propose performing a backward sweep.

\begin{figure}[h!]
    \centering
    \includegraphics[width=\columnwidth]{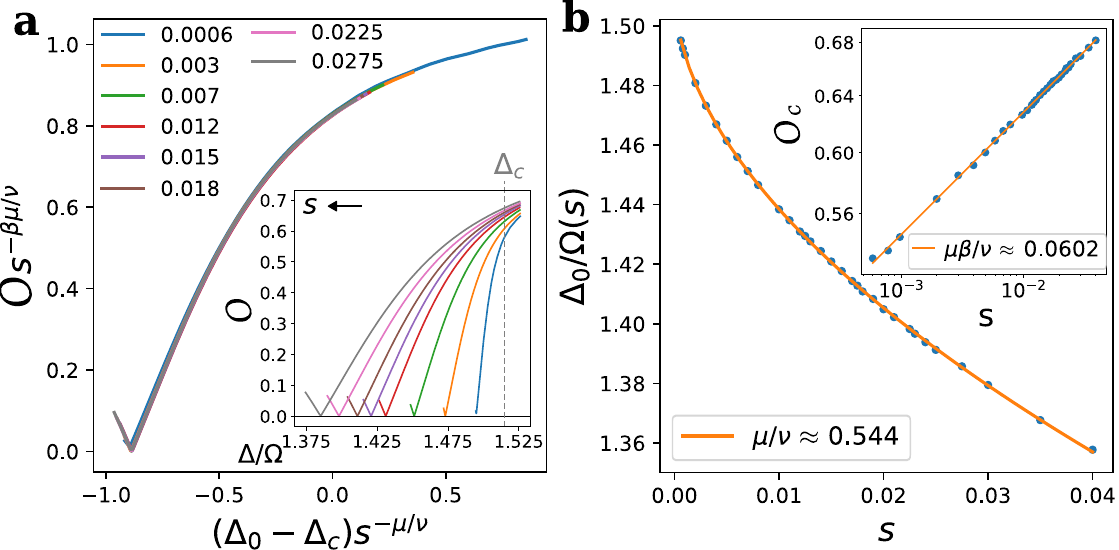}
    \caption{Example of the finite-time scaling out of period-3 phase. (a) Order parameter as a function of the sweep rate after the re-scaling indicated at each axis. The inset shows the same curves before the re-scaling. (b) Location of the point $\Delta_0$, where the order parameter vanishes as a function of sweep rate. Inset: Scaling of the order parameter $O$ at the critical point $\Delta_c$ as a function of sweep rate $s$. Orange lines are fits with Eq.(\ref{eq:fts1}) and (\ref{eq:fts2}).}
    \label{fig:example_fts}
\end{figure}

{\bf Sweeping from the ordered to the disordered phase} - in the direction opposite to the KZ mechanism - allows to study a relaxation of the order. The faster is the sweeping rate the longer it takes for the order parameter\cite{SM} to disappear after crossing the transition, as illustrated by the inset of Fig.\ref{fig:example_fts}(a). We extract the ratio $\mu/\nu$ by fitting the location where the order parameter vanishes with Eq.(\ref{eq:fts1}), as shown in  Fig.\ref{fig:example_fts}(b).
Furthermore, we extract $\beta\mu/\nu$ in the inset by fitting the order parameter at the critical point with Eq.(\ref{eq:fts2}). 

\begin{figure}[h!]
    \centering
    \includegraphics[width=0.5\textwidth]{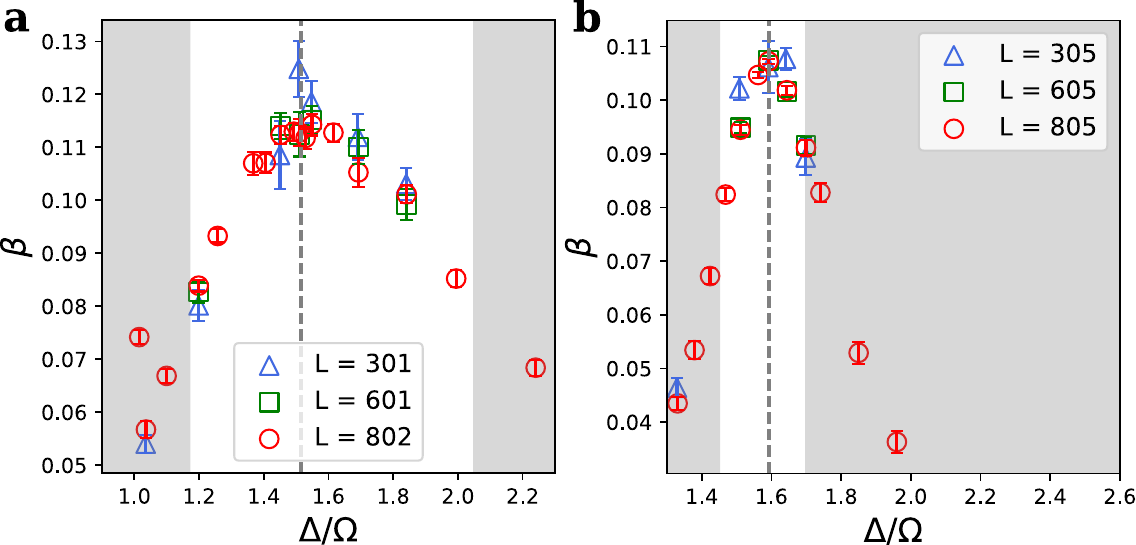}
    \caption{Critical exponent $\beta$ measured with finite-time scaling of the order parameter while sweeping from the ordered (a) period-3 and (b) period-4 phases to the disordered one. In both cases, the conformal points (dashed lines) correspond to the pronounced peaks in $\beta$. Shaded regions indicate the floating phases.}
    \label{fig:blockade_fts_distribution}
\end{figure}

Combining the results from the two fits, we extract the critical exponent $\beta$ along various cuts from period-3 and period-4 phases. The results are summarized in Fig.\ref{fig:blockade_fts_distribution}(a) and (b) correspondingly.
Similar to the Kibble-Zurek critical exponent $\mu$, we find that $\beta$ takes its maximal values at the conformal points. However, the sharp peaks in $\beta$ allow identifying the location of the critical point with significantly smaller uncertainty.

It is important to keep in mind that the critical exponent $\beta$ is well understood only for the two conformal points. At the three-state Potts point, it takes the universal value $\beta=1/9$. Our numerical result $\beta \approx 0.112$ agrees within $\sim 2\%$ with this value.
As long as the transition to the disordered phase is direct and the order parameter goes to zero at the transition, the exponent $\beta$ can be formally defined. However, its value is unknown and might be affected by the domain wall tension\cite{huse1982domain}. Indeed, what we observe is that $\beta$ is not universal and varies along the chiral transition, decaying away from the conformal point. This fully agrees with previous numerical results on the chiral clock model where the exponent $\beta$ for chiral transitions has been extracted\cite{huang2019nonequilibrium}.

Ashkin-Teller critical theory defines the family of universality classes with the exponents $\frac{1}{12}\leq \beta\leq\frac{1}{8}$ and $\frac{2}{3}\leq \nu\leq 1$ satisfying $d=\beta/\nu=1/8$\cite{PhysRevB.91.165129}.
Our numerical result $\beta \approx 0.1073$ belongs to the corresponding interval. Equilibrium simulations have reported an Ashkin-Teller point with $\nu\approx0.78$\cite{chepiga2021kibble}, implying $\beta\approx0.098$. This value agrees with our out-of-equilibrium result within $10\%$.
Similar to the period-3 case, we see that the value of $\beta$ is not universal along the $p=4$ chiral transition.

{\bf Combination of forward and backwards sweeping} opens a way to extract all critical exponents, including the dynamical critical exponent $z$  presented in Fig.\ref{fig:blockade_kz_distribution_overlap}(a) and (c): at the two conformal points our numerical results match the CFT value $z=1$; away from these points $z$ increases, in agreement with previous studies  \cite{keesling2019quantum,chepiga2021kibble}. We also extract the critical exponent $\alpha$. Unlike other critical exponents, $\alpha$ does not change significantly along the chiral transitions. This agrees with the idea that along chiral transitions, $\alpha$ keeps the value it takes at the conformal point\cite{albertini,Baxter1989,Cardy1993}.

\begin{figure}[h!]
     \centering
        \includegraphics[width=0.5\textwidth]{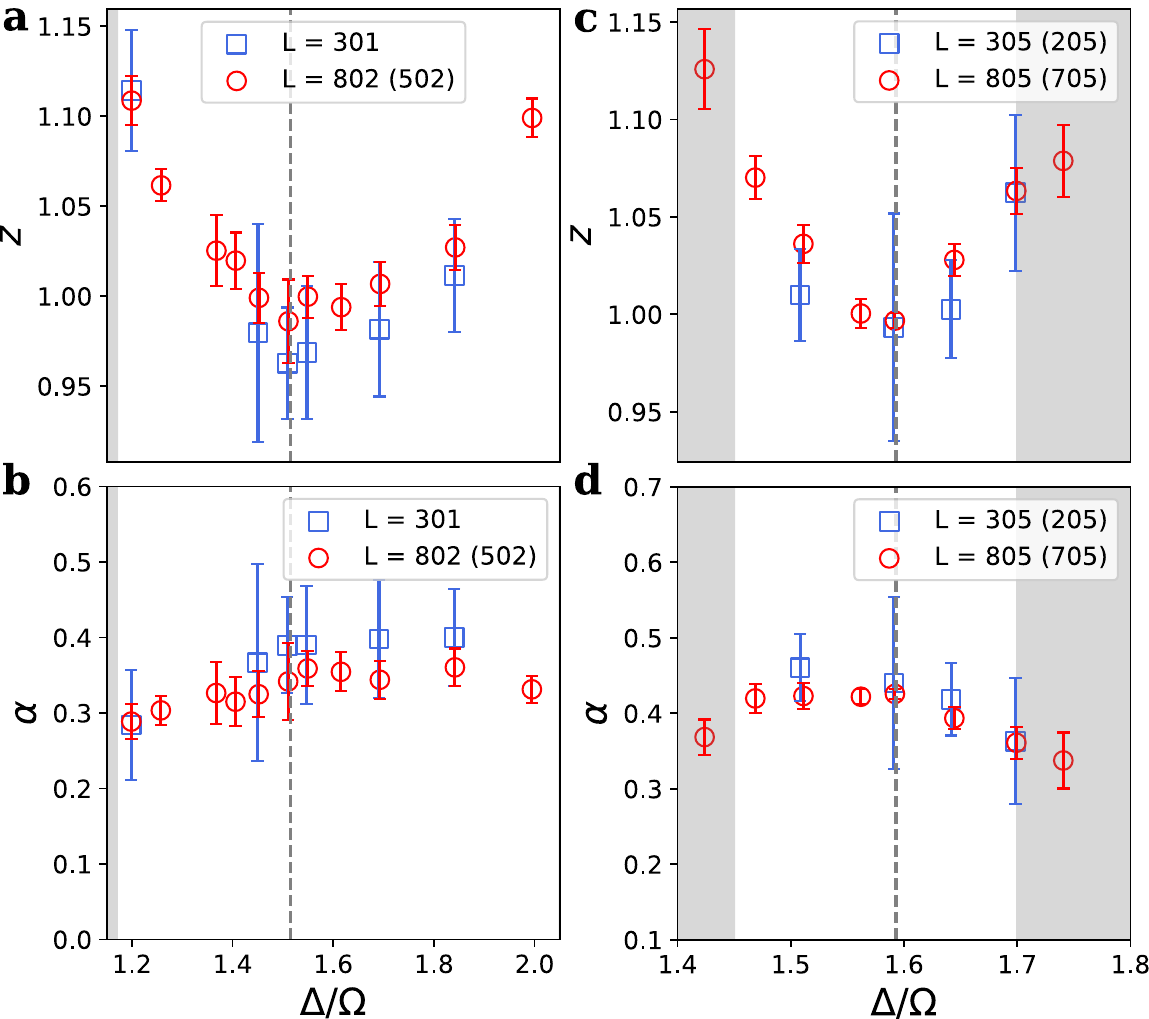}
        \caption{ Dynamical critical exponent $z$ and specific heat critical exponent $\alpha$ computed across various cuts across transitions to the (a)-(b) period-3 and (c)-(d) period-4 phases. (a),(c) Our results agree with $z=1$ at the two conformal points (dashed lines) and suggest that $z>1$ when the transitions are chiral. (b),(d) Numerically extracted values of $\alpha$ are in good agreement with CFT predictions $\alpha=1/3$ for three-state Potts and $\alpha\approx0.44$ for the Ashkin-Teller point with $\nu\approx0.78$.  We show the results for small (blue) and large (red) system sizes; when $L$ used in KZ and in finite-time scaling are different, the former is indicated in brackets.  Shaded regions state for the floating phases.  }
        \label{fig:blockade_kz_distribution_overlap}
\end{figure}

{\bf Discussion}. In the present manuscript, we have shown how quantum phase transitions can be fully characterized by combining Kibble-Zurek dynamics with finite-time scaling of the order parameter. We demonstrated that the appearance of the intermediate floating phase can be identified by comparing Kibble-Zurek dynamics in Rydberg arrays with different numbers of atoms. Our approach relies on the standard Kibble-Zurek protocol and, by contrast to previous proposals, does not require measurements of the correlation length, incommensurate wave-vector\cite{chepiga2021kibble} or structure factor\cite{zhang2024probing} near or inside the critical region. 

We have also shown that by sweeping from the ordered to disordered phase and keeping track of the order parameter, the location of the conformal points can be accurately identified with the critical exponent $\beta$. 
Although this method requires several runs terminating at different distances to the transitions, the ordered phases are less sensitive to noise, and the sampling might require only a few runs. In order to prepare a high-quality ordered state and to ensure an identical starting point for all samples, one can use a recently developed light-shift method\cite{pham2022coherent}. Interestingly enough, when the location of the transitions is known, the KZ dynamics can be combined with measurements of the remaining order at the transition to extract the scaling dimension of the corresponding operator.

Our predictions for blockade models remain valid for the model Eq.\ref{eq:tail} with van der Waals interaction. Performing a finite-time scaling of the period-3 order we observe a sharp peak in $\beta$ \cite{SM} that agrees with previously identified location of the conformal point\cite{maceira2022conformal}. At the same point, the KZ exponent $\mu$ takes its maximum.
The interval of the $p=4$ chiral transition in this model is very narrow\cite{maceira2022conformal}, making it extremely challenging for dynamical studies. However, the protocols developed here are generic and can be applied to multi-component systems, where the extent of the chiral transition can be controlled\cite{chepiga2024tunable}.

By analyzing the critical exponent $\beta$ in Fig.\ref{fig:blockade_fts_distribution} we made an interesting observation:  the value of $\beta$ when chiral transition turns into a floating phase seems to be roughly the same on both sides of the conformal points. With uncertainly in our data and the location of the Lifshitz points, we cannot exclude a simple coincidence, however, the possibility of some critical values of $\beta$ beyond which the chiral transition is unstable might deserve further investigation.

{\bf Acknowledgments.} 
 NC acknowledges useful discussions with Hannes Bernien, Frederic Mila and Rui-Zhen Huang. This research has been supported by Delft Technology Fellowship.  
  Numerical simulations have been performed at the DelftBlue HPC and at the Dutch national e-infrastructure with the support of the SURF Cooperative.

\bibliographystyle{unsrt}
\bibliography{bibliography,comments}
\pagebreak

\section*{Supplemental material to "Resolving chiral transitions in Rydberg arrays with quantum Kibble-Zurek mechanism and finite-time scaling"}
\section{Blockade Models}
\subsection{Implementation details}

This section explains how Rydberg blockade can be encoded into one-dimensional (1D) tensor network. Note that various blockade ranges give access to different slices of the phase diagram, as explained in Ref.\onlinecite{chepiga2021kibble}. 
In this paper, we are focusing on transitions out of period-3 and period-4 phases. The tip of the period-3 phase,  where the conformal three-state Potts critical point and two intervals of chiral transitions are realized, can be addressed with the $r=1$ blockade model\cite{fendley,chepiga2019floating,maceira2022conformal}. The tip of the period-4 phase, encompassing the Ashkin-Teller conformal point and two intervals of the $p=4$ chiral transition, can be addressed with the $r=2$\cite{chepiga2021kibble}.

\begin{figure}[h!]
    \centering
    \includegraphics[width = \columnwidth]{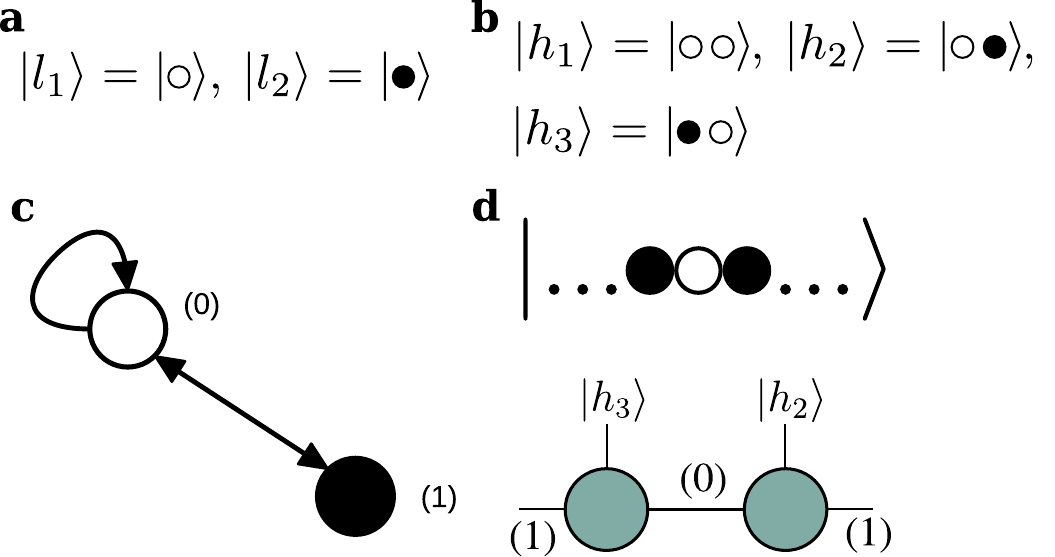}
 \caption{Mapping of the $r=1$ blockade model onto a model preserving the block diagonal structure of tensors. (a) Local Hilbert space $\ket{l_i}$ of the original model. (b) New local Hilbert space spanned over two consecutive sites. (c) Fusion graph for the recursive construction of the environment (both, left and right): starting with empty site (0), another empty site can always be added, ending up with label (0). Additionally, an occupied site can be added, leading to the label (1). On the other side, starting with label (1), only an empty site can be added, which results in label (0). (e) Example of the label assignment in MPS representation on two consecutive tensors written for the selected state.}
     \label{fig:1newbasis}
\end{figure}

To fully profit from the reduced Hilbert space, a new basis was taken, wherein each element of the new basis is composed of a pair of adjacent local basis elements, where the last element of each tensor overlaps with the first site on the following tensor. In other words, we span the local physical degrees of freedom of each individual tensor over two consecutive atoms. Using the occupation of the atom shared between two nearest tensors as a quantum label of the auxiliary bond that connects them, we can bring the network into a block-diagonal form. The latter drastically reduces the computational complexity. Figure \ref{fig:1newbasis} illustrates how the $r=1$ blockade basis is constructed.

The bulk Hamiltonian in the new $\ket{h_i}$ basis is:
\begin{multline}
    h_{i} = -\frac{\Omega}{2}\left(\tilde{a}_{i}\tilde{b}_{i+1} + \textnormal{h.c.}\right)
   - \Delta\tilde q_i + V_2 \tilde q_i \tilde p_{i+1}. 
\end{multline}

The term $\tilde{a}_{i}\tilde{b}_{i+1} + \textnormal{h.c.}$ represents the first term in Eq.(5a) of the main text. This term flips an atom between ground and Rydberg states.
It comprises three local sites, with $\tilde a \ket{h_1} = \ket{h_2}$, and $\tilde b \ket{h_1} = \ket{h_3}$, and $0$ otherwise. The operators $\tilde q_i$ and $\tilde p_i$ are local density operators for the left and right sites in each pair, returning non-zero values only for $\tilde q \ket{h_3} = \ket{h_3}$ and $\tilde p \ket{h2} = \ket{h_2}$ correspondingly. As a result, the next-nearest-neighbor interaction $V_2n_in_{i+2}$ transforms into a nearest-neighbor interaction $V_2\tilde q_i \tilde p_{i+1}$. In order to match the original Hamiltonian of Eq.(5), this effective model must be carefully adapted close to the boundary. \\

The explicit implementation of $r=2$ blockade is conceptually very similar. We span the local degrees of freedom over three consecutive atoms and use the quantum state of the two atoms shared by nearest tensors as quantum labels for their common auxiliary leg. The bulk Hamiltonian has the form:
\begin{multline}
    h_i = -\frac{\Omega}{2}\left(\tilde{a}_{i}\tilde{b}_{i+1}\tilde{c}_{i+2} 
    + \textnormal{h.c.}\right)
    - \Delta\tilde q_i + V_3\tilde q_i \tilde p_{i+1}. 
\end{multline}

In this case, the term $\tilde{a}_{i}\tilde{b}_{i+1}\tilde{c}_{i+2} + \textnormal{h.c.}$ represents the first term in Eq.(5a) of the main text, which flips an atom between ground and Rydberg states.
The only non-zero elements are $\tilde a \ket{h_1} = \ket{h_2}$,  $\tilde b \ket{h_1} = \ket{h_3}$ and $\tilde c \ket{h_1} = \ket{h_4}$. The terms $\tilde q_i$ and $\tilde p_i$  represent the density of the local sites $i$ and $i+2$ correspondingly, with only a single non-zero entree $\tilde q \ket{h_4} = \ket{h_4}$ and $p_i\ket{h_1} = \ket{h_1}$. The interaction $V_3n_in_{i+3}$ transforms into a nearest-neighbor interaction $V_2\tilde q_i \tilde p_{i+1}$ under the new basis. Similarly to the $r=1$ case, this Hamiltonian has to be adapted close to the edges to capture the boundary terms.\\

\begin{figure}[h!]
    \centering
    \includegraphics[width = \columnwidth]{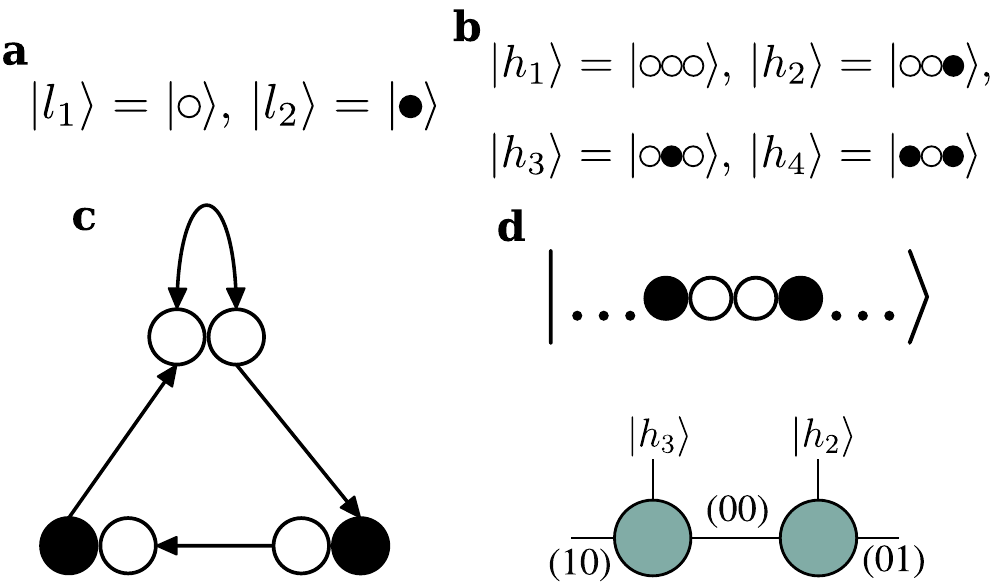}
    \caption{Mapping of the $r=2$ blockade model onto a model preserving the block diagonal structure of tensors. (a) Local Hilbert space of the original model $\ket{l_i}$. (b) New local Hilbert space spanned over three consecutive sites. (c) Fusion graph for the recursive construction of the left environment (for the right environment, the direction of arrows should be reverted). (e) Example of the label assignment in MPS representation on two consecutive tensors written for the selected state.}
\end{figure}

\subsection{Ground State Calculations}

The initial state defined at time $t = 0$ is a ground state at a given point in the phase diagram sufficiently far from the transition. This point is located either in the disordered phase (corresponding to the starting point for the Kibble-Zurek mechanism) or in the ordered period-3 or period-4 phases (the starting points for the backward sweeps for the finite-time scaling of the order parameter).

The ground state was determined with imaginary time-evolving block decimation (TEBD) for the two blockade models. We used second-order Trotter decomposition.
Maximal bond dimension was kept at $D=300$, and singular values below $\chi > 10^{-6}$ were truncated. Convergence criteria were based on the order parameter, with calculations considered converged when the relative variation of the order parameter was smaller than $ 10^{-9}$.

Deep inside the ordered phase, the correlation length is very small, and the energy cost of the domain wall formation is relatively cheap. Therefore, there are many low-lying excited states above the ground states. When the system is close to the classical limit, and the entanglement is low, the TEBD is often stuck at such states with multiple domain walls. To circumvent this issue, we take as a starting guess a classical state (with $D=1$) that resembles the expected pattern of occupied and empty sites and then perform 4 sweeps to converge for a given quantum point inside the ordered phase.  

\subsection{Simulation of dynamics}

Simulations of dynamics in the blockade model were performed using a second-order time-evolving block decimation algorithm (TEBD). For the $r=1$ blockade, a two-site Trotterization was applied, while for the $r=2$ blockade, a three-site Trotterization was used. The time step was maintained at $\delta t = 0.1$, and the maximum bond dimension and singular value cutoff were set to $D = 300$ and $\chi > 10^{-6}$ correspondingly.

\begin{figure}[h!]
    \centering
    \includegraphics[width =\columnwidth]{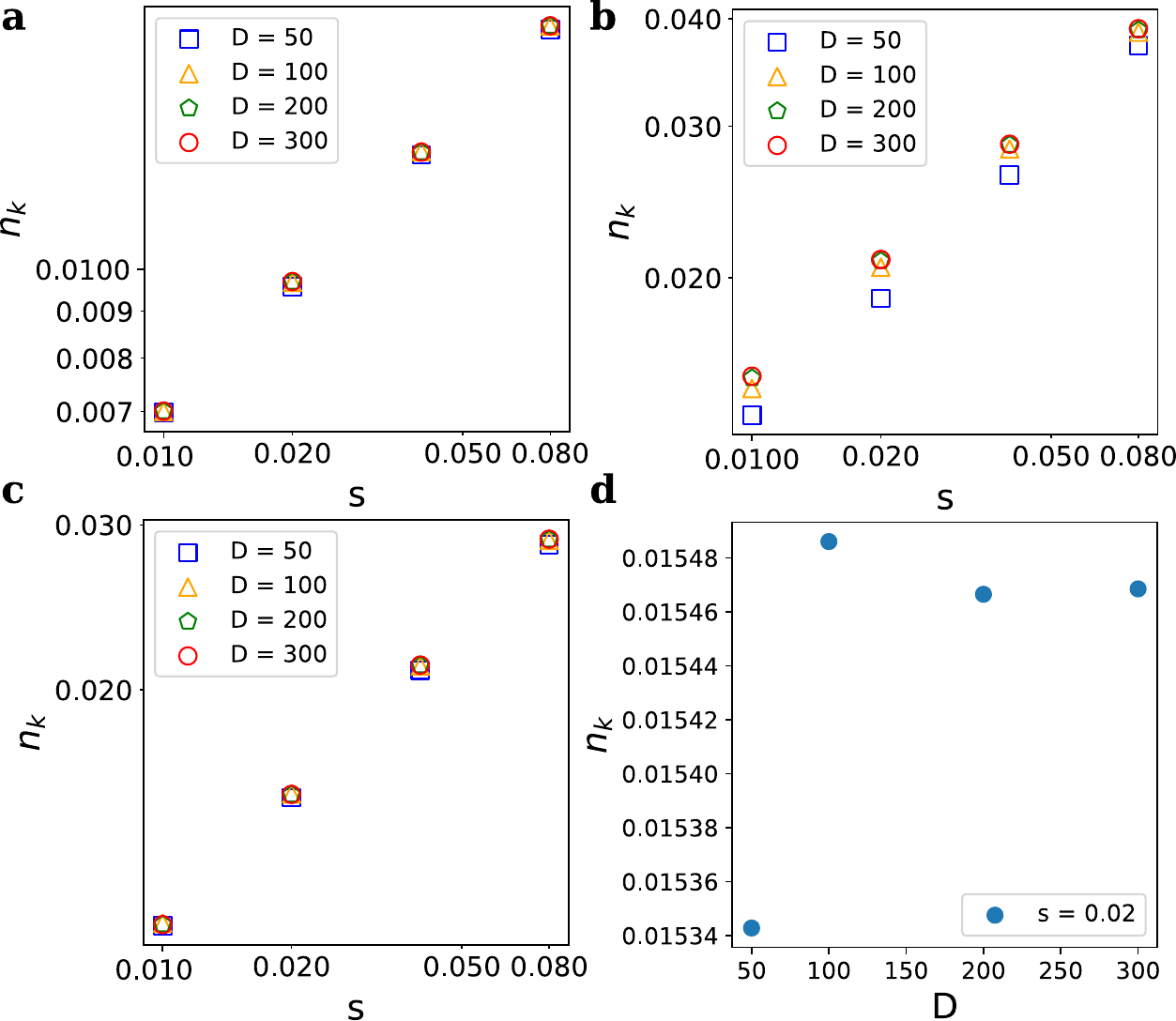}
    \caption{Density of kinks as a function of sweep rate in a $\log-\log$ scale for various maximal bond dimensions $D$ measured after crossing (a) the three-state Potts transition; (b) the Ashkin-Teller transition; (c) the chiral transition. (d) Density of kinks formed across the chiral transition for a given sweep rate as a function of bond dimension $D$. }
    \label{fig:convergence}
\end{figure}

Fig.\ref{fig:convergence} shows convergence with respect to bond dimension for Kibble-Zurek(KZ) dynamics across transitions of various types. In Fig.\ref{fig:convergence}(a) and (b), we present the extracted density of kinks formed across the three-state Potts and for the Ashkin-Teller point correspondingly as a function of sweep rates for bond dimension ranging from $D=50$ to $D=300$. It is clear from the figure that the finite-bond dimension effect is stronger for the Ashkin-Teller point, but in both cases, the results for the two largest bond dimensions are indistinguishable. 

Fig.\ref{fig:convergence}(c) shows the density of kinks formed while crossing the $p=3$ chiral transition. Fig.\ref{fig:convergence}(d) shows the convergence of the extracted density of kinks for a given sweep rate $s$ as a function of bond dimension $D$.

\section{Kink operators}
\begin{figure}[h!]
    \centering
    \includegraphics[width=0.9\columnwidth]{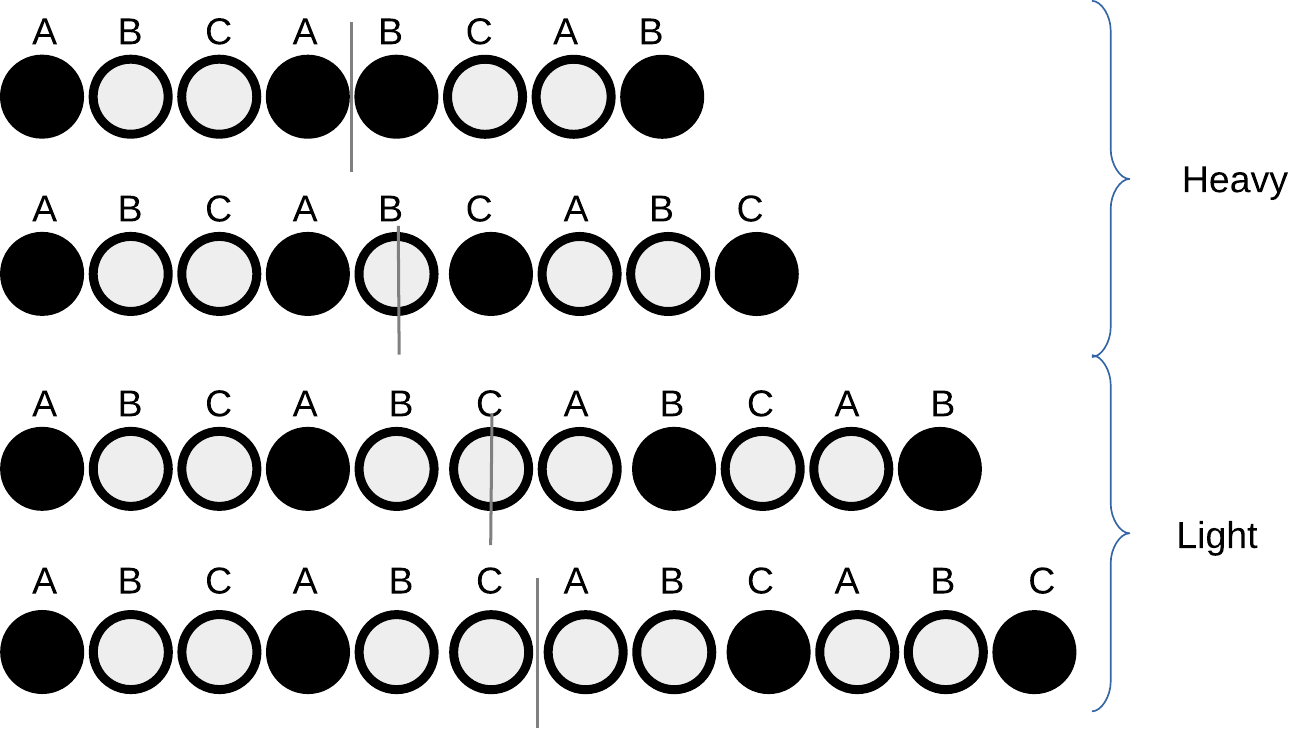}
    \caption{Types of domain walls in the period 3 phase. There are three types of domains, depending on the label $A$, $B$ or $C$ on top of the black dot.}
    \label{fig:domain_walls}
\end{figure}

Kinks can be counted in two different ways. One way is counting the number of domain walls. Fig.\ref{fig:domain_walls} depicts the types of domain walls that can appear in period-3 as described in Ref.\onlinecite{huse1982domain}. An alternative method for counting kinks is using an operator that counts the absence of an ordered state. This \textit{no-order} operator would be $(1 - \fullcircle\emptycircle\emptycircle - \emptycircle\emptycircle\fullcircle - \emptycircle\fullcircle\emptycircle)$ in the period-3 case, and quantifies a kink every time the periodicity of the phase is not followed. These two methods of counting kinks are not always equivalent. For instance, the first method would count one kink in the state $\fullcircle\emptycircle\emptycircle\fullcircle\fullcircle\emptycircle\emptycircle\fullcircle$, while the second method would count two kinks. Additionally, the first method would not count any kink in the state
 $\fullcircle\emptycircle\emptycircle\emptycircle\emptycircle\emptycircle\fullcircle$, since the type of domain of the chain does not change, while the second method would count three kinks. 

\begin{figure}[h!]
    \centering
    \includegraphics[width=0.8\columnwidth]{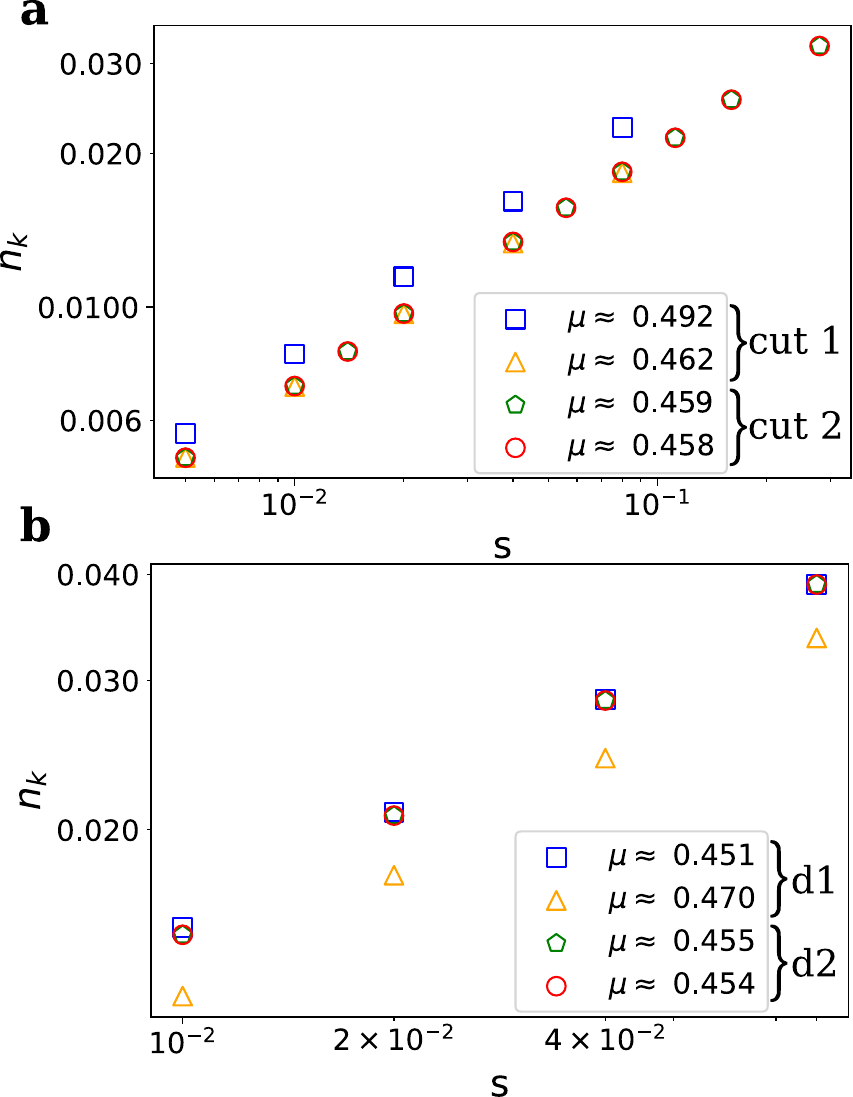}
    \caption{KZ scaling for the density kinks measured as domain walls (orange triangles and red circles) or with a \textit{no-order} operator (blue squares and green pentagons) for (a) Two different cuts through the Potts point for the $r=1$ blockade model (b) Two final distances from the AT points for the $r = 2$ blockade model.}
    \label{fig:kinks}
\end{figure}

In Fig.\ref{fig:kinks}, we compare the KZ scaling using the two abovementioned counting methods. Fig.\ref{fig:kinks}(a) compares two different cuts through the Potts point for a $r = 1$ blockade, while Fig.\ref{fig:kinks}(b) compares two final distances from the AT point in the $r=2$ blockade model. In all cases, blue squares and green pentagons represent the density of kinks measured with the no-order operator while orange triangles and red circles represent the density of domain walls as described in Fig.\ref{fig:domain_walls}

For the $r = 1$ blockade model, data points representing the density of domain walls for the two different cuts and the density kinks using the no-order operator for cut 2 overlap and are close to the theoretical value $\mu = 0.454$. In contrast, data points representing the density of kinks for the no-order operator are off and far from $0.454$. However, for $r = 2$ the opposite situation occurs. Data points representing the density of kinks measured with the no-order operator for the two different final distances d1 (short) and d2 (long) and the density of domain walls for d2 overlap and are close to the theoretical value $\mu = 0.44$, while the density of domain walls for d1 is off and far from $0.44$. Similar effects were found for different cuts and final distances from the critical point, where convergence always occurred faster for the density of domain walls in the $r=1$ blockade and for the density of kinks calculated with the no-order operator for the $r=2$ blockade.

We believe that the discrepancy between the two methods comes from the rare appearance of unlikely events - the state $\fullcircle\emptycircle\emptycircle\fullcircle\fullcircle\emptycircle\emptycircle\fullcircle$ is forbidden by the blockade, while the state $\fullcircle\emptycircle\emptycircle\emptycircle\emptycircle\emptycircle\fullcircle$ is energetically more costly than any other type of domain walls sketched in Fig.\ref{fig:domain_walls}. However sweeping along certain trajectories, for example, those along which incommensurability is non-monotonous\cite{chepiga2019floating,chepiga2021kibble} the probability of these events might increase. In our research we compared different trajectories, different start and end points and pick ups the kink operators that is the most robust: domain wall operators for period-3 and no-order operator for period-4. 

\section{Finite-size effects in Kibble-Zurek mechanism}

\begin{figure}[h!]
    \centering
    \includegraphics[width =\columnwidth]{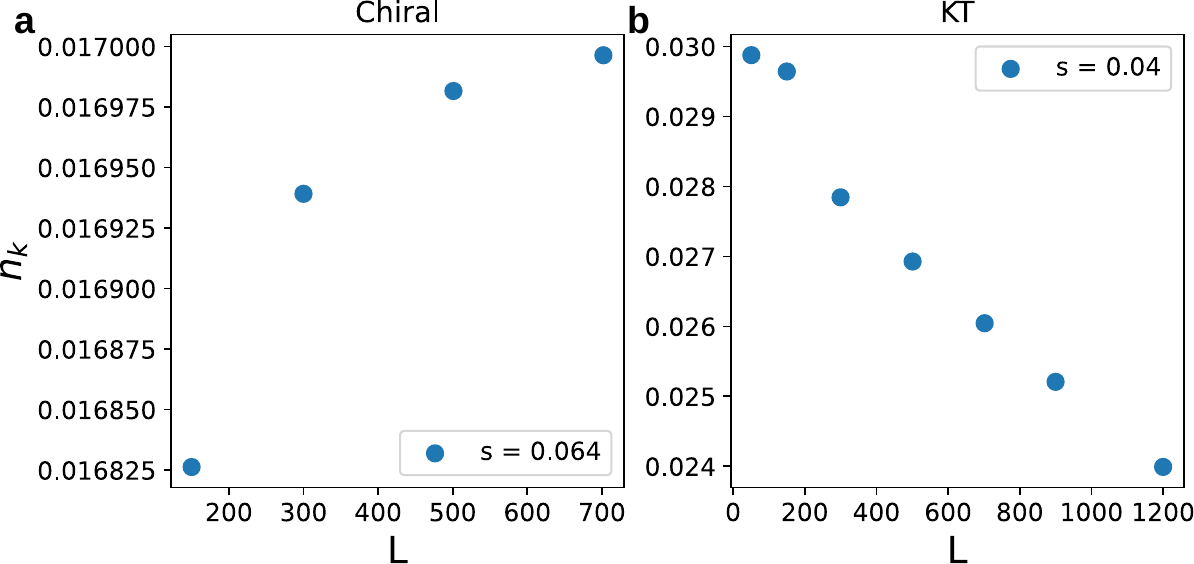}
    \caption{Finite-size effect of the density of kinks formed by sweeping from the disordered phase to the period-3 phase across (a) the direct chiral transition and (b) the intermediate floating phase separated from the disordered phase by the Kosterlitz-Thouless (KT) transition. The value of sweep rate $s$ is indicated at each panel. }
    \label{fig:fss_r1}
\end{figure}

In the main text, we argue that in the Kibble-Zurek mechanism, the intermediate floating phase can be distinguished from the direct transition by tracking the finite-size effects in the density of kinks formed by sweeping through the criticality. Fig.\ref{fig:fss_r1} shows the density of kinks as a function of the system size for the two cuts presented in Fig.(1) of the main text. Fig.\ref{fig:fss_r1}(a) depicts a cut through the chiral transition. Through this cut, the density of kinks increases with the system size $L$. Fig.\ref{fig:fss_r1}(b) shows the cut through the floating phase, which is mainly governed by the Kosterlitz-Thouless (KT) transition between the disordered and the floating phases. In contrast to the direct transition, the density of kinks systematically decreases with the system size. 

\begin{figure}[h!]
    \centering
    \includegraphics[width = \columnwidth]{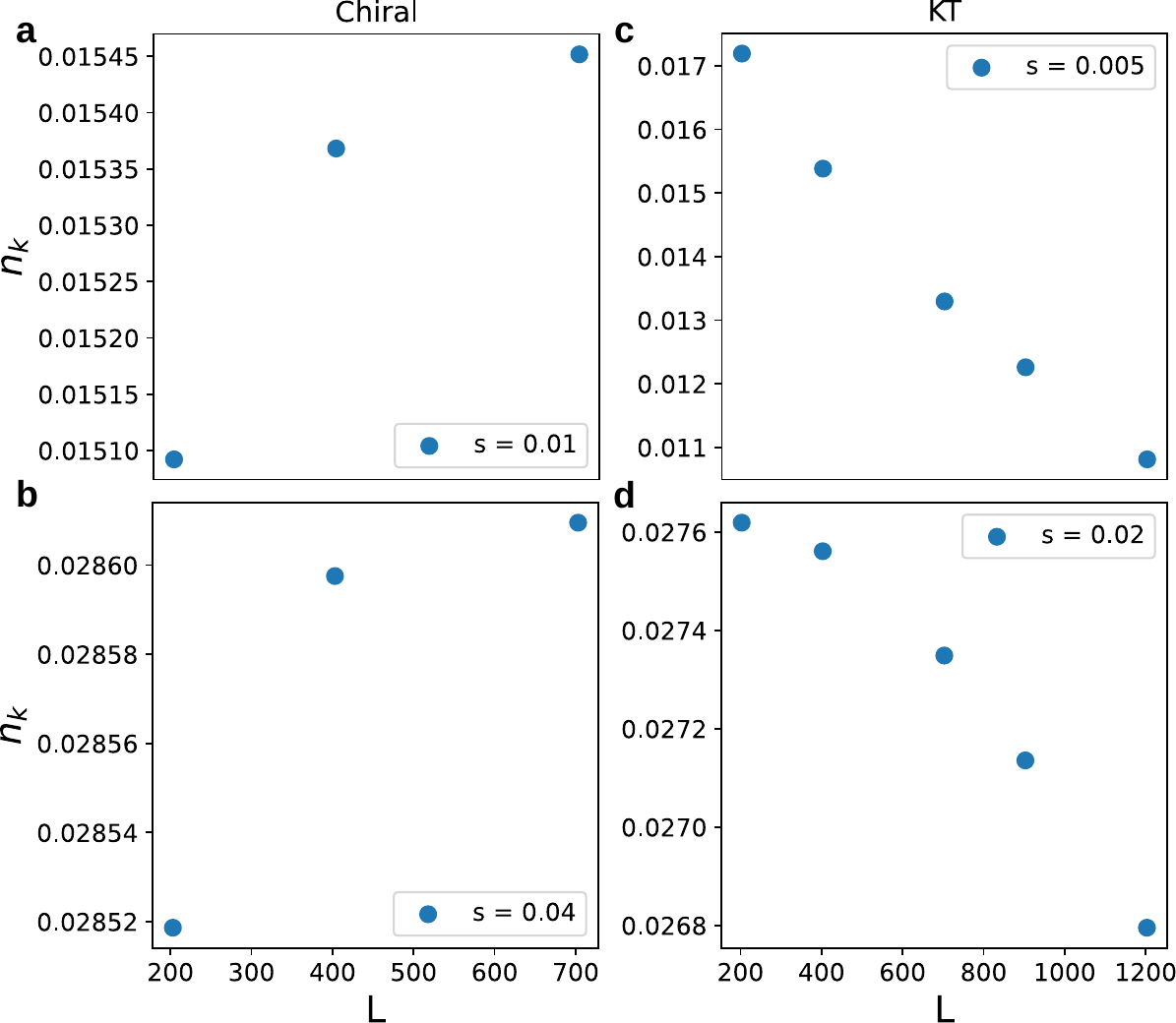}
    \caption{Finite-size effect of the density of kinks formed by sweeping from the disordered phase to the period-4 phase across (a),(b) the direct chiral transition and (c),(d) the intermediate floating phase separated from a disordered phase by the Kosterlitz-Thouless (KT) transition. The trajectories for (a) and (b) and for (c) and (d) are identical. }
    \label{fig:fss_r2}
\end{figure}

The Kibble-Zurek mechanism across the transitions into the period-4 phase with a $r=2$ blockade model shows qualitatively similar finite-size effects as in the period-3 case. The results for period-4 are summarized in Fig.\ref{fig:fss_r2}.

\section{Order parameter}

In this section, we define the order parameter associated with the gaped period-3 and period-4 phases with broken translation symmetry. As an order parameter, we use the local amplitude of the local density that we average over the whole finite-size chain.
 Explicitly, for the period-3 case, we use: 
 \begin{equation}
     O = \frac{1}{L-2}\sum_{i=1}^{L-2}\textnormal{max}_i\left(\abs{n_i - n_{i+1}},\abs{n_i - n_{i+2}}\right)
 \end{equation} 
and for the period-4:
  \begin{equation}
  O = \frac{1}{L-3}\sum_{i=1}^{L-3}\textnormal{max}_i\left(\abs{n_i - n_{i+1}},\abs{n_i - n_{i+2}}, \abs{n_i - n_{i+3}}\right)
  \end{equation}

\section{Finite-size effect in the finite-time scaling}

\begin{figure}[h!]
         \includegraphics[width=\columnwidth]{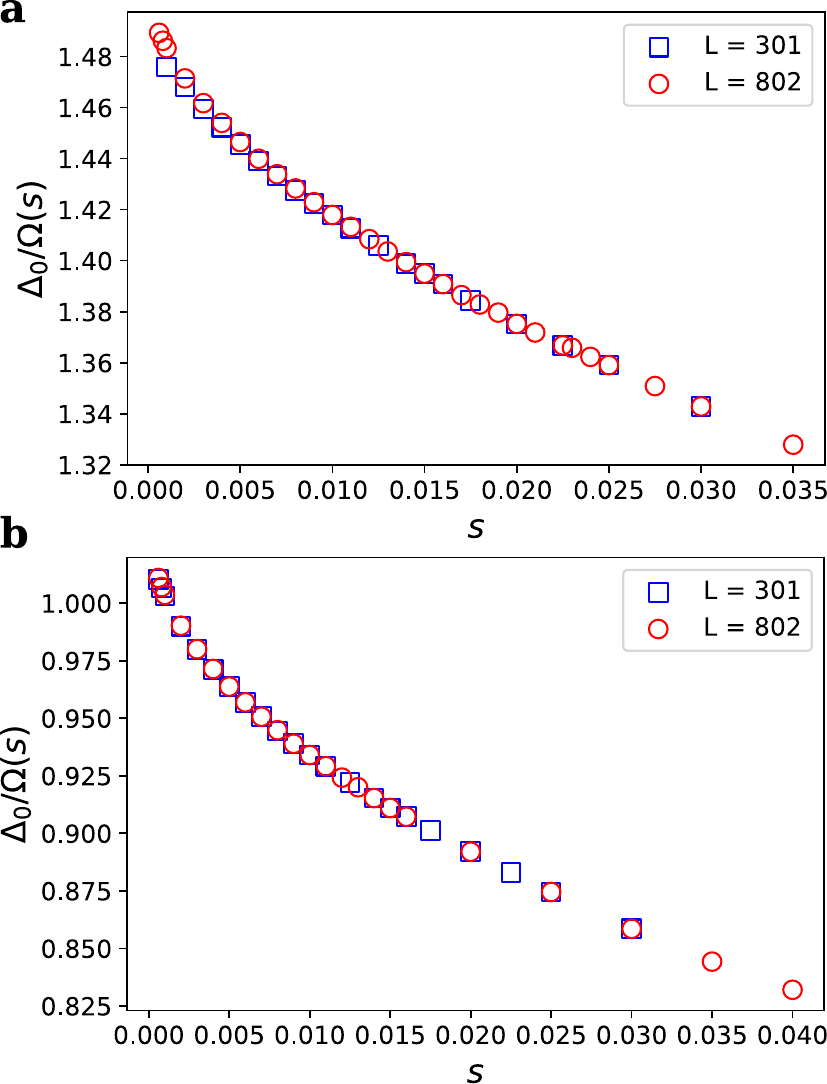}
        \caption{Comparison of the system size effect in the finite-time scaling for (a) a direct transition and (b) an intermediate floating phase for two different system sizes. In both figures, the difference is visible only for very small sweep rates.}
        \label{fig:FTS_system_size}
\end{figure}

Figure \ref{fig:FTS_system_size} compares the finite-time scaling (FTS) for two different system sizes. For both system sizes,  direct transition (see Fig.\ref{fig:FTS_system_size}(a)) and sweep through a floating phase (see Fig.\ref{fig:FTS_system_size}(b)), we observe an almost perfect overlap except for the slowest sweep rates.

\section{Derived critical exponents}

\begin{figure}[h!]
    \centering
    \includegraphics[width = \columnwidth]{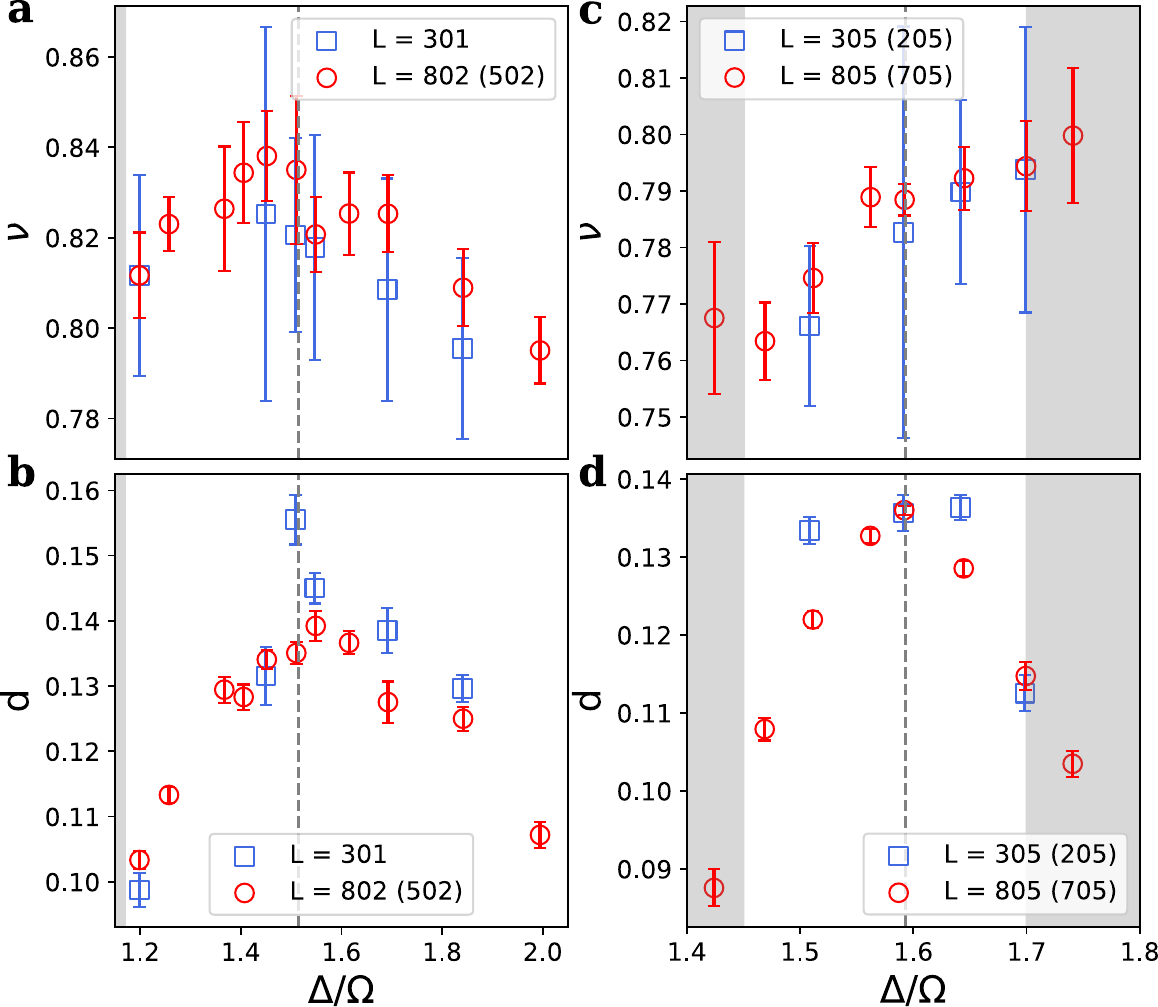}
    \caption{Correlation length critical exponent $\nu$ (a,c) and scaling dimension of the order parameter $d=\beta/\nu$ (b,d) extracted by combining the Kibble-Zurek mechanism with finite-time scaling across the transition (a-b) into period-3  and (c-d) into period-4  phases. }
    \label{fig:nud}
\end{figure}

In this section, we present additional results of the critical exponents that can be derived from the Kibble-Zurek mechanism and the finite-time scaling of the order parameter. In particular, we show the correlation length critical exponent $\nu$ across various cuts into the period-3 (see Fig.\ref{fig:nud}(a)) and into the period-4 (see Fig.\ref{fig:nud}(c)) phases. In both cases, the critical exponent $\nu$ of the chiral transition stays within $\approx5-6\%$ of the value at the corresponding conformal point. Our results for the period-3 case (see Fig.\ref{fig:nud}(a)) agree with the results on the chiral clock model reporting the decrease of $\nu$ away from the Potts point\cite{huang2019nonequilibrium}. Curiously enough, for the period-4 case, we see that on one side of the transition, $\nu$ might increase. It would be interesting to clarify the behaviour of $\nu$ with a more accurate systematic estimate of $\nu$ across the period-4 transition. 

In addition, we extract the scaling dimension $d=\beta/\nu$ of the operator that defines the order parameter. These results are summarized in Fig.\ref{fig:nud}(b),(d).

\section{Rydberg model with $1/r^6$ interaction}

In this section, we present the results obtained for the model with van der Waals $1/r^6$ potential as defined in Eq.(4) of the main text.

\subsection{Technical details}

For the two-site density matrix renormalization group calculations\cite{dmrg1,dmrg3}, the decaying $1/r^6$ van der Waals interaction was expressed as a sum of $11$ exponentials \cite{pirvu2010matrix, schollwock2011density}, i.e., $1/r^6 = \sum_{i=1}^{11} c_i\lambda_i^r$. The coefficients $c_i$ and exponents $\lambda_i$ were determined by minimizing the cost function defined as:
\begin{equation}
\sum_{r=1}^L\abs{\frac{1}{r^6}- \sum_{i=1}^{11}c_i\lambda_i^r}.
\end{equation}
This optimization process followed the method described in \cite{pirvu2010matrix}. Convergence was declared when variations in the energy per site were $\Delta E < 10^{-8}$.

To match the ordered phase and boundary conditions, the system sizes were chosen in the form $L = N p + 1$, where $N$ is an integer, and $p$ is the periodicity of the ordered phase.

\subsection{Simulation of dynamics}
For the Rydberg model with $1/r^6$ van der Waals interactions, time evolution was simulated using the Time-Dependent Variational Principle (TDVP) \cite{haegeman2011time, haegeman2016unifying}. The long-range interactions were approximated using a sum of 7 exponentials; the maximum error in this approximation was $\sim 10^{-10}$, with a cost function  $\sim 1.3 \times 10^{-19}$. Different $\delta t$  for various sweep rates and critical points were tested during 24h simulations. The values of $\delta t$ that gave the fastest simulations were taken for surrounding sweep rates and critical points. Truncation criteria were maintained at $\chi > 10^{-6}$, and the maximum bond dimension was set to $D = 400$.\\

\subsection{ Ising transition}

We benchmark our method with Ising transition\cite{keesling2019quantum,rader2019floating} into period-2 phase for which there are theory predictions for all critical exponents ($\nu = 1$, $z = 1$, $\mu = 0.5$).

\begin{figure}[h!]
    \centering
    \includegraphics[width = 0.9\columnwidth]{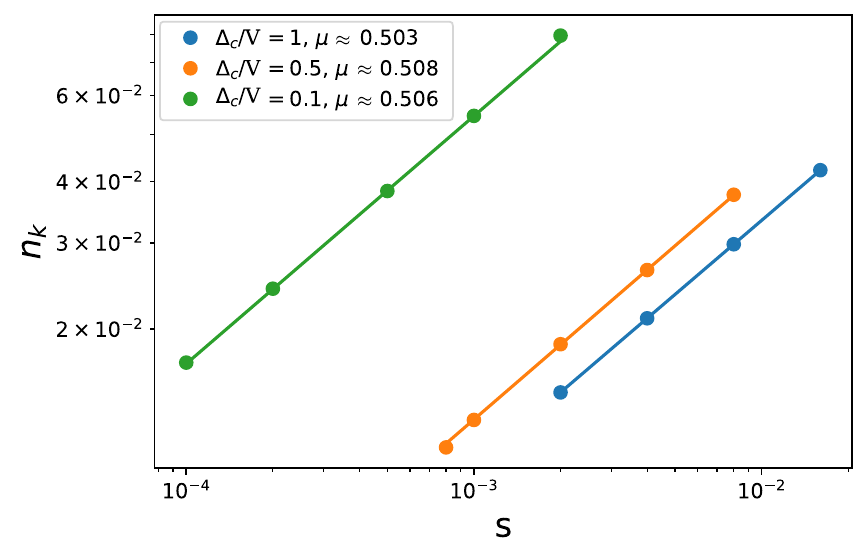}
    \caption{Scaling of the density of kinks $n_k$ with the sweep rate for a disordered to period $p = 2$ phase transition in the Rydberg model at different critical values of detuning $\Delta_c$. The transition belongs to the Ising universality class characterized by a Kibble-Zurek exponent $\mu = 0.5$. The extracted value of $\mu$ is in excellent agreement with theoretical predictions}
    \label{fig:rydberg_ising}
\end{figure}

Figure \ref{fig:rydberg_ising} show scaling of the density of kinks and extracted Kibble-Zurek critical exponent $\mu$ for three different cuts across the Ising transition - the agreement is always within $2\%$. 

\subsection{Transitions into the period-3 phase}

\begin{figure}[h!]
    \centering
    \includegraphics[width = \columnwidth]{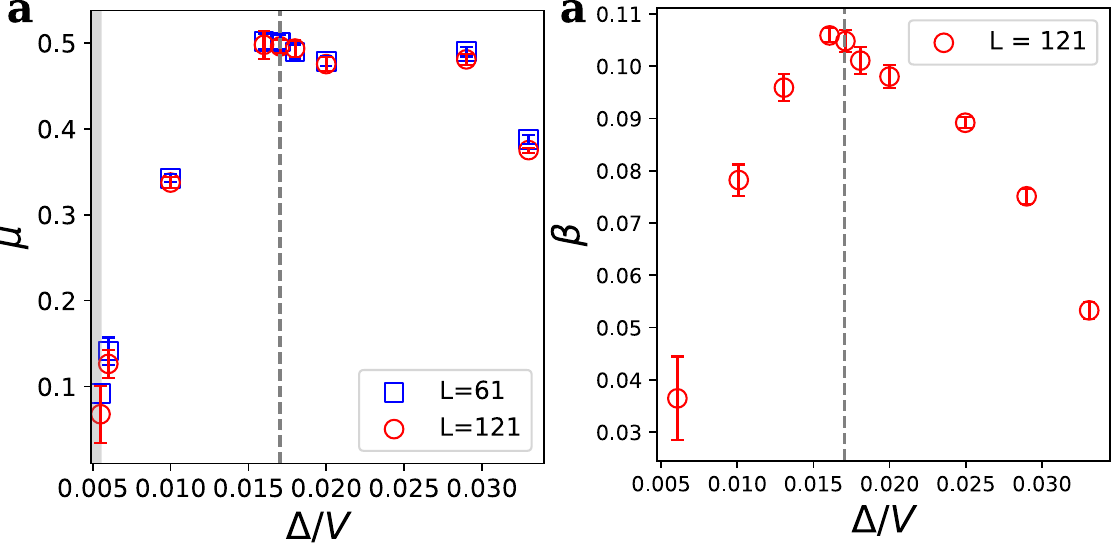}
    \caption{Critical exponents (a) $\mu$ and (b) $\beta$ computed across various cuts for the period 3 Rydberg model with $1/r^6$ interaction using KZ and FTS, respectively. Both $\mu$ and $\beta$ have a dome shape that peaks at the conformal three-state Potts point. This peak is more pronounced for $\beta$, which agrees with theory prediction $\beta = 1/9$ for the three-state Potts point within $5\%$. Numerical results for $\mu$ agree within $8\%$ with CFT predictions $\mu\approx0.454$ for the three-state Potts point.}
    \label{fig:rydberg_chiral}
\end{figure}

By sweeping from the disordered to the period-3 phase, we extract the critical exponent $\mu$ for the model defined by Eq.(4) of the main text. The results are summarized in Fig.\ref{fig:rydberg_chiral}(a).  The point where the critical exponent $\mu$ takes maximal value is in the good agreement with previously identified location of the three-state Potts point. The discrepancy is probably due to the reported strong finite-size effect on the location of the transition. Interestingly, below the Potts point, $\mu$ decays vary slowly in agreement with an extended interval of the chiral transition.

We perform a finite-time scaling and extract the critical exponent $\beta$ by keeping track of the order parameter in the backward sweep. Similar to the blockade model, we see a sharp peak in $\beta$ around the conformal point. At the three-state Potts point, the extracted value of $\beta$ agrees with the theory prediction $\beta=1/9$ within $5\%$.

\end{document}